\documentclass[final,5p,times]{elsarticle}
\usepackage{amssymb}
\usepackage{amsmath}  % Greek letter
\usepackage{graphicx} %Graph
\usepackage{caption}
\usepackage{epstopdf}
\usepackage{subfigure}
\usepackage{color}
\usepackage{xcolor}
\usepackage{booktabs}

\journal{xxx}

\begin{document}

\bibliographystyle{elsarticle-num}
\begin{frontmatter}



\title{Monthly electricity consumption forecasting by the fruit fly optimization algorithm enhanced Holt-Winters smoothing method}

%% use optional labels to link authors explicitly to addresses:o
\author[label1]{Weiheng Jiang}
\ead{whjiang@cqu.edu.cn}
\author[label1]{Xiaogang Wu}
%%\ead{xiaogangwu@cqu.edu.cn}
\author[label2]{Yi Gong}
\author[label1]{Wanxin Yu}
\author[label1]{Xinhui Zhong}
\address[label1]{Chongqing University, Chongqing 400044, China.}
\address[label2]{Southern University of Science and Technology, Shenzhen 518055, China}
%%\author{Xiaogang Wu, Weiheng Jiang}

%%\address{College of Communication Engineering, Chongqing University, Chongqing 400044, China.}

\begin{abstract}
The electricity consumption forecasting is a critical component of the intelligent power system. And accurate monthly electricity consumption forecasting, as one of the the medium and long term electricity consumption forecasting problems, plays an important role in dispatching and management for electric power systems. Although there are many studies for this problem, large sample data set is generally required to obtain higher prediction accuracy, and the prediction performance become worse when only a little data is available. However, in practical, mostly we experience the problem of insufficient sample data and how to accurately forecast the monthly electricity consumption with limited sample data is a challenge task. The Holt-Winters exponential smoothing method often used to forecast periodic series due to low demand for training data and high accuracy for forecasting. In this paper, based on Holt-Winters exponential smoothing method, we propose a hybrid forecasting model named FOA-MHW. The main idea is that, we use fruit fly optimization algorithm to select smoothing parameters for Holt-Winters exponential smoothing method. Besides, electricity consumption data of a city in China is used to comprehensively evaluate the forecasting performance of the proposed model. The results indicate that our model can significantly improve the accuracy of monthly electricity consumption forecasting even in the case that only a small number of training data is available.
\end{abstract}


\end{frontmatter}


\section{Introduction}
\label{Introduction}
Electricity consumption forecasting is a critical component for the modern management of the electric power systems, which has attracted more and more attentions from both the academic and the industry society \cite{DEBNATH2018297}. And highly-accurate mid-term and long-term electricity consumption forecasting plays an important role in dispatching and planning of electric power systems. However, due to the fact that the mid-term or long-term electricity consumption has complex and non-linear relationships with lots of external factors, such as the political environment, economic policy, human activities, irregular behaviors and other non-linear factors \cite{Li2013A}, which then makes it is difficult to forecast accurately. In order to improve the accuracy of mid-term and long-term electricity consumption forecasting, various prediction methods and models are proposed these years.

From the perspective of forecasting model, the mid-term or long-term electricity consumption forecasting methods can be classified into two types, i.e., the stand-alone based model and the hybrid model \cite{DEBNATH2018297} \cite{Shao2017A}. In addition, based on the used techniques, the stand-alone model is further divided into three categories, i.e., statistical model \cite{Tratar2016The} \cite{Aydin2014Modeling}, computation intelligence (CI) model \cite{KIRAN201293} \cite{Zhang2018A}, and the mathematical programming (MP) based model \cite{forouzanfar2010modeling}. While for the hybrid methods, they are further divided into four categories, i.e., the statistical-statistical \cite{tan2010day}, statistical-CI \cite{da2019bottom}, CI-CI \cite{Ju2013Application} \cite{Chen2018Forecasting}, and statistical-MP methods \cite{forouzanfar2010modeling}. The stand-alone methods can also be divided into linear and nonlinear categories \cite{Hernandez2014A}.

In particular, for the stand-alone model, lots of work have been done over these different categories of models. Firstly, for the statistical model, the main methods include exponential smoothing method \cite{Tratar2016The} \cite{RENDONSANCHEZ2019916},  regression analysis method \cite{Aydin2014Modeling} \cite{Tsekouras2007A}, and the time series method \cite{DEOLIVEIRA2018776} \cite{Nawaz2014Modelling}. In detail, Tsekouras et. al \cite{Tsekouras2007A} presented a non-linear multi-variable regression method for midterm energy forecasting of the power systems in annually \cite{HE2019565}. The author of \cite{Nawaz2014Modelling} estimated Pakistan's electricity demand by applying STAR (Smooth Transition Auto-Regressive) model. Secondly, the computational intelligence methods include metaheuristic method \cite{KIRAN201293} \cite{Amjadi2010Estimation}, machine learning method \cite{Zhang2018A} \cite{Taylor2002Neural}, knowledge-based method \cite{Kandil2002Long} \cite{TANG20191144} and the uncertainty method \cite{Wang2012Optimization} \cite{Ali2016Long}. More specifically, the regressive convolution neural network (RCNN) is used to extract features from data and then the regressive support vector machine (SVR) trained with features is adopted to predict the electricity consumption, which achieves a low predicting error in \cite{Zhang2018A}. In addition, instead of point forecasting, based on fuzzy Bayesian theory and expert prediction, a novel long-term probability forecasting model is proposed to predict the Chinese per-capita electricity consumption (PEC) and its variation interval over the period 2010-2030 \cite{TANG20191144}.

Comparing with the stand-alone model, the hybrid model is more popularly studied due to its excellent representation ability of nonlinear factors and random factors. Concretely, Ju et. al employed chaotic gravitational search algorithm to determine the three free parameters in the support vector regression (SVR) model, and which can significantly improve the performance of the SVR \cite{Ju2013Application}. In \cite{Chen2018Forecasting}, a novel approach for monthly electricity demands forecasting by the integration of both the wavelet transform and the neuro-fuzzy system is proposed, and the results confirm that this model can provide an accurate forecasting. He et. al proposed a method of probability density forecasting based on least absolute shrinkage and selection operator-quantile regression neural network (LASSO-QRNN), and the prediction accuracy of the proposed model is evaluated through the empirical analyses with the Guangdong province dataset in China and the California dataset in the United States \cite{HE2019565}. In \cite{da2019bottom}, a methodology that combines the bottom-up approach with hierarchical linear models for long term electricity consumption forecasting of a particular industrial sector considering energy efficiency scenarios is proposed, and the model was used to generate long term point and probability distribution forecasts for the period
ranging from 2015 until 2050.

Till now, though lots of work have been done for the mid-term electricity consumption forecasting and different models have been presented. It is noted that, all these proposed models can only handle the case with large number of sample data, and have low forecasting accuracy if we only have a small number of training sample data, especially for the computational intelligence method and the machine learning based method. However, in practical, as we know that, in China, the development of the informatization for the electrical companies in some areas only beginning recently, thus the aggregate historical electricity consumption data is limited. Thus, how to forecast the electricity consumption accurately with insufficient sample data is a challenge problem.

In this paper, a new model based on the integration of the fruit fly optimization algorithm (FOA) and the Holt-Winters exponential smoothing (HW) is proposed for the mid-term electricity consumption forecasting, to handle the situation that we have only a small number of aggregated historical electricity consumption data. On the one hand, HW is known as a common approach and used to forecast the seasonal time series \cite{Holt2004Forecasting} \cite{Winters1960Forecasting}. However, the performance of HW is mainly depended on the selection of appropriate smoothing parameters. On the other hand, we know that the FOA is an excellent tool in pursuing the global optimization for the parameter optimization problems \cite{Pan2012A} \cite{Li2013A}. Therefore, herein, we introduce the FOA to assist the smoothing parameter selection for the HW and which formulates the proposed FOA-MHW model. To the best knowledge of the authors, this is the first enhanced HW model, and the real data set based test shows that it can obtain incredible performance even in the situation of insufficient sample data.

The rest of this paper is organized as follows: Section 2 introduces the Holt-Winters exponential smoothing method and the FOA, and then a hybrid forecasting model constructed by the Holt-Winters exponential smoothing method and FOA is proposed and discussed. In Section 3, the sample data set used in this paper is illustrated, and then based on this sample data set, the performance of the proposed model are evaluated and compared under different length of training data sets. At last, we conclude this paper in Section 4.

\section{Methodology of FOA-MHW model}
In this section, at first, we introduce the Holt-Winters (HW) exponential smoothing model, and then the theory
 of fruit fly optimization algorithm is presented. At last, the process of the proposed FOA-MHW model is described in details.
\label{Model}
\subsection{Holt-Winters exponential smoothing method}
\label{ModelHW}
Exponential smoothing is an important time series forecasting method, its basic idea is to preprocess the original data to eliminate the randomness in the time series, and improve the importance of the recent data in predicting the collected data. The processed data is called "smoothing value", and then the forecasting model is constructed according to the smoothing value, and the future target value is forecasted by the model \cite{ForeastBook}. However, the simple exponential smoothing method can't overcome the randomness of time series enough, to this end, the Holt-Winters exponential smoothing method is then developed by Holt and Winters \cite{Holt2004Forecasting} \cite{Winters1960Forecasting}. Previous analysis shows that, even with a small number of the sample data, the Holt-Winters exponential smoothing method still can achieve a good forecasting result \cite{ForeastBook}. While the Holt-Winters exponential smoothing method can be classified into four types: 1) Multiplicative model (MHW), 2) Additive model (AHW), 3) Modified model (MoHW), 4) Extended model (EHW). The differences among these models come from the methods used in calculating the seasonal indices \cite{Tratar2016The}. In which, the MHW is the most popular Holt-Winters exponential smoothing method dues to the fact that, the calculation method of seasonal indices is more suitable for the actual situation. Hence, we choose MHW as our primary forecasting model.

The multiplicative Holt-Winters (MHW) model forecasts trend, stationarity and seasonal components of the time series at the same time. The basic form of the Holt-Winters exponential smoothing forecasting model is as follows:
 \begin{equation}
  \hat{y}_{t+m} = (T_t + b_tm)S_{t+m-L}
  \label{eq:1}
 \end{equation}
In which, $\hat{y}_{t+m}$ means the $m$th forecasting data, $t$ is the length of training data and herein, it denotes the length of the period. Furthermore, the forecasting model includes three components of time series: stationarity $T_t$, trend $b_t$ and seasonality $S_t$. $T_t$ is used to correct time series to rule out seasonal factors in time series, $b_t$ is used to correct trend values to eliminate seasonal interference in the time series, and $S_t$ is used to forecast seasonal indices to exclude random interference \cite{ForeastBook}. The recursive process of the three variable are as follows.
\begin{equation}
  T_t = \alpha \frac{y_t}{S_{t-L}} + (1- \alpha)(T_{t-1}+b_{t-1})
  \label{eq:2}
 \end{equation}

\begin{equation}
  b_t = \beta (T_t - T_{t-1}) + (1 - \beta)b_{t-1}
  \label{eq:3}
 \end{equation}

 \begin{equation}
  S_t = \gamma \frac{y_t}{T_t} + (1 - \gamma)S_{t-L}
  \label{eq:4}
 \end{equation}
Herein, $\alpha$, $\beta$ and $\gamma$ are three smoothing parameters of the Holt-Winters model, and all of them take value from $[0,1]$. In addition, $y_t$ denotes the observation data. Based on the analysis of the first order exponential smoothing method \cite{ForeastBook}, it is found that the larger value of $\alpha$, $\beta$ and $\gamma$, the smaller influence of the long-term historical electricity quantity on the prediction, and vice versa. Hence, these smoothing parameters play a critical role in the final predicting results and how to determine the value of these parameters is the focus of this paper.

The initialization method of the equation (2) - (4) is as follows.
\begin{equation}
  T_0 = \frac{1}{n} \sum_{t=1}^n y_t
  \label{eq:5}
\end{equation}

\begin{equation}
  S_{0k} = \frac{\bar{y_k}}{T_0}
  \label{eq:6}
\end{equation}

\begin{equation}
  b_0 = 0
  \label{eq:6}
\end{equation}

where,
\begin{equation}
  \bar{y_k} = \frac{1}{n/L}\sum_{t=1 \wedge t = jL+k}^n y_t, (k = 1, ..., L, j = 1,2, ..., n/L-1)
  \label{eq:6}
\end{equation}
Herein, $T_0$ is the average value of the training data, $S_{0k}$ ($k = 1, ..., L$) is the initial value of the seasonal index, and the initial value of $b_t$ is set to 0.

\subsection{Fruit fly optimization algorithm}
\begin{figure}[htb]
\centerline{\includegraphics[scale=0.65]{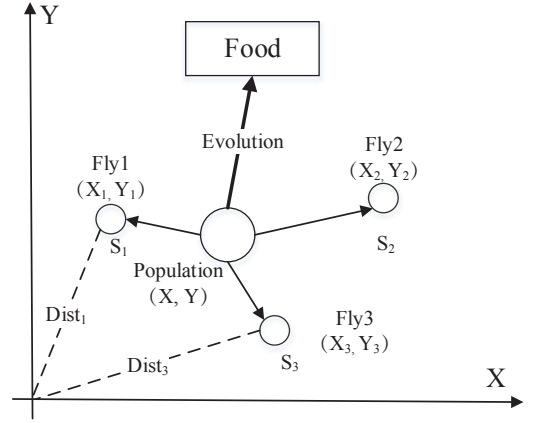}}
\caption{$~$The evolution process of fruit fly population \cite{Li2013A}.}
\label{FOA}
\end{figure}

Fruit fly optimization algorithm (FOA) is an intelligent optimization algorithm developed by Pan in \cite{Pan2012A}. And it is often used to search the global optimal solution for parameter optimization problem based on the food finding behavior of the fruit fly. The fruit fly itself is superior to other species in sensing and perception, especially in osphresis and vision. The flow chart of the evolution of fruit fly population is shown in Figure \ref{FOA}, and the details of the algorithm are summarized as follows \cite{Cao2016Support}.

\begin{itemize}
  \item[(1)] Parameter setting: The population size ($sizepop$), the maximum iteration number ($maxgen$), and the random flight range ($FR$), the initial fruit fly swarm location ($x_0$,$y_0$).
  \item[(2)] Initialization: Give the random direction and distance for the search of food using osphresis by an individual fruit fly.
  \begin{equation}
  X_i = x_0 + rand(FR),
  \label{eq:1}
  \end{equation}
  \begin{equation}
  Y_i = y_0 + rand(FR),
  \label{eq:2}
  \end{equation}
  \item[(3)] Calculate the smell concentration judgment value: Calculate the distance between the fruit fly individual to the origin ($Dist_i$), then calculate the smell concentration judgment value ($S_i$).
  \begin{equation}
  Dist_i = \sqrt{X_i^2+Y_i^2},
  \label{eq:3}
  \end{equation}
  \begin{equation}
  S_i = 1/Dist_i;
  \label{eq:4}
  \end{equation}
  \item[(4)] Calculate the smell concentration: Feed the $S_i$ into the fitness function, thus obtain the smell concentration ($Smell_i$) of the individual fruit fly location, i.e., the equation (5).
  \begin{equation}
  Smell_i = Fuction(S_i),
  \label{eq:5}
  \end{equation}
  \item[(5)] Find the best individual: Find out the fruit fly with maximal smell concentration (finding the maximal value) among the fruit fly swarm.
  \begin{equation}
  [bestSmell, bestIndex] = max(Smell_i),
  \label{eq:6}
  \end{equation}
  Where the $bestSmell$ represent the highest smell concentration among the current fruit fly swarm, and the $bestIndex$ represents the fruit fly which has the highest smell concentration among the current fruit fly swarm.
  \item[(6)] Fruit fly swarm movement: Select the best fruit fly individual $bestIndex$, keep the best direction, and at this moment, the fruit fly swarm will use vision to fly towards that best location.
  \begin{equation}
  smellBest = bestSmell
  \label{eq:7}
  \end{equation}
  \begin{equation}
  x_0 = X_{bestIndex}
  \label{eq:8}
  \end{equation}
  \begin{equation}
  y_0 = Y_{bestIndex}
  \label{eq:9}
  \end{equation}
  \item[(7)] Population evolution: Perform the iterative optimization and repeat the implementations from step 2 to step 6,  when the smell concentration is not better than the previous iteration results any more or the number of iteration reaches the $maxgen$, the algorithm is ended.
\end{itemize}

\subsection{Fruit fly optimization algorithm for parameter selection of the Holt-Winters exponential smoothing method}
\label{FOA-HW}
As we have mentioned in section \ref{ModelHW}, the forecasting performance of the Holt-Winters exponential smoothing method is mainly determined by the value of the three smoothing parameters, i.e., $\alpha$, $\beta$, and $\gamma$. Therefore, in this paper, the FOA is used to search the value of these smoothing parameters for the Holt-Winters exponential smoothing model and which will significantly improve the accuracy of the monthly electricity consumption forecasting.
\begin{figure}[htb]
\centerline{\includegraphics[scale=0.45]{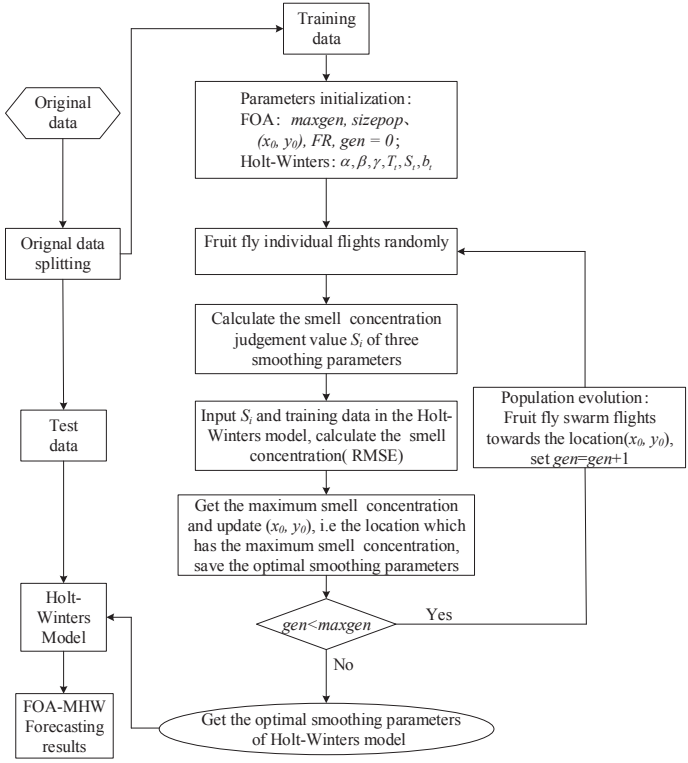}}
\caption{$~$The flowchart of FOA selecting smoothing parameters for the Holt-Winters exponential smoothing model.}
\label{FOAHW}
\end{figure}

The flowchart of the FOA based smoothing parameter selecting for the Holt-Winters exponential smoothing model (abbreviated as FOA-MHW) is presented in Figure \ref{FOAHW}. And the details of the algorithm are summarized as follows.

\begin{itemize}
  \item[(1)] Data sets splitting: Divide the sample data set into training data set and test data set. In addition, the last period of training data set is defined as validation data set, and which is used to conduct the optimal smoothing parameter selection in the FOA for Holt-Winters exponential smoothing model.
  \item[(2)] Parameter initialization: Initialize the population size ($sizepop$), the maximum iteration number ($maxgen$), the random flight range ($FX^\alpha$, $FY^\alpha$), ($FX^\beta$, $FY^\beta$), ($FX^\gamma$, $FY^\gamma$) and the initial fruit fly swarm location ($x_0^\alpha$, $y_0^\alpha$), ($x_0^\beta$, $y_0^\beta$) and ($x_0^\gamma$,$y_0^\gamma$).

      Where $(x_0^\alpha$, $y_0^\alpha)$ is the initial fruit fly swarm location of parameter $\alpha$, $(x_0^\beta$, $y_0^\beta)$ denotes the initial fruit fly swarm location of parameter $\beta$ and $(x_0^\gamma$, $y_0^\gamma)$ characterizes the initial fruit fly swarm location of parameter $\gamma$. Hence, there are three fruit flies searching their optimal smoothing parameters for the Holt-Winters exponential smoothing model.
  \item[(3)] Fruit fly swarm flight: Each of the three fruit fly randomly searches of the food at a predetermined distance, and each of fruit fly's location of these three fruit fly swarms are $(X_i^\alpha, Y_i^\alpha)$, $(X_i^\beta, Y_i^\beta)$ and $(X_i^\gamma, Y_i^\gamma)$, respectively.

      Where,
      \begin{equation}
      X_i^\alpha = x_0^\alpha + FX^\alpha, Y_i^\alpha = y_0^\alpha + FY^\alpha
      \label{eq:9}
      \end{equation}
      \begin{equation}
      X_i^\beta = x_0^\beta + FX^\beta, Y_i^\beta = y_0^\beta + FY^\beta
      \label{eq:9}
      \end{equation}
      \begin{equation}
      X_i^\gamma = x_0^\gamma + FX^\gamma, Y_i^\gamma = y_0^\gamma + FY^\gamma
      \label{eq:9}
      \end{equation}
  \item[(4)] Calculate the smell concentration judgment value: the smell concentration judgment value ($S_i$) of fruit fly is calculated as follows.
      \begin{equation}
      S_i^\alpha = \alpha = \frac{1}{D_i^\alpha}
      \label{eq:9}
      \end{equation}
      \begin{equation}
      S_i^\beta = \beta = \frac{1}{D_i^\beta}
      \label{eq:9}
      \end{equation}
      \begin{equation}
      S_i^\gamma = \gamma = \frac{1}{D_i^\gamma}
      \label{eq:9}
      \end{equation}

      Where,
      \begin{equation}
      D_i^\alpha = 1/\sqrt{X_i^{\alpha2} + Y_i^{\alpha2}}
      \label{eq:9}
      \end{equation}
      \begin{equation}
      D_i^\beta = 1/\sqrt{X_i^{\beta2} + Y_i^{\beta2}}
      \label{eq:9}
      \end{equation}
      \begin{equation}
      D_i^\gamma = 1/\sqrt{X_i^{\gamma2} + Y_i^{\gamma2}}
      \label{eq:9}
      \end{equation}
      Herein, $D_i^\alpha$, $D_i^\beta$ and $D_i^\gamma$ denote the distance between the fruit fly individual and the origin. We feed the smell concentration judgment value (use as $\alpha$, $\beta$ and $\gamma$) and training data into the Holt-Winters model to obtain a forecasting result, then we calculate the root mean square error (RMSE) between the forecasting results and the validation data set.
      \item[(5)] Calculate the smell concentration: we choose the RMSE as the smell concentration ($Smell_i$), and it is calculated as follows.
      \begin{equation}
      RMSE = \sqrt{\sum_m (y_m-\hat{y_m})^2}
      \label{eq:9}
      \end{equation}
      Where $y_m$ is the data in validation data set, and $\hat{y_m}$ is the data of forecasting results.
      \item[(6)] Find the best individual: Find out the three fruit flies with maximal smell concentration (finding the maximal value) among the fruit fly swarm, as equation (14).
      \item[(7)] Fruit fly swarm movement: Select the best three fruit fly individuals $bestIndex$, and save the smell concentration judgment value. Keep the best direction, and at this moment, the fruit fly swarm will use vision to fly towards that best location, as follows.
      \begin{equation}
      x_0^\alpha = X_{bestIndex}^\alpha, y_0^\alpha = Y_{bestIndex}^\alpha,
      \label{eq:9}
      \end{equation}
      \begin{equation}
      x_0^\beta = X_{bestIndex}^\beta, y_0^\beta = Y_{bestIndex}^\beta,
      \label{eq:9}
      \end{equation}
      \begin{equation}
      x_0^\gamma = X_{bestIndex}^\gamma, y_0^\gamma = Y_{bestIndex}^\gamma,
      \label{eq:9}
      \end{equation}
      \item[(8)] Population evolution: Repeat the implementations of step 3 to step 7, when the smell concentration is not better than the previous iteration results any more or the number of iteration reaches the $maxgen$, then we get the best smell concentration judgment value, i.e. the optimal smoothing parameters $\alpha_{opt}$, $\beta_{opt}$ and $\gamma_{opt}$ and go to step 9.
      \item[(9)] Forecasting: Feed the optimal smoothing parameters and the training data set into the Holt-Winters Model to obtain the final forecasting result.
 \end{itemize}

\section{Numerical example}
\label{Example}
Two examples are selected to verify the effectiveness of the proposed FOA-MHW model in this paper, which are the monthly electricity consumption forecasting of a southern city in China and the telecommunications and television industry in this city. In order to testify the performance of the proposed algorithms, three typical algorithms are introduced herein as the benchmarks, they are the seasonal index (SI) model \cite{WANG2012109}, the MHW model with default parameters \cite{ForeastBook} and the GASVR model \cite{OYEHAN201885}. In addition, to verify the performance gain of our FOA-MHW model under small sample data set, we evaluate the performance of FOA-MHW model over different lengthes of training data set. The performance metric used herein is the mean absolute percentage error (MAPE) and it is defined as below.
\begin{equation}
MAPE = \frac{1}{N}\sum_{t=1}^N \left| \frac{\hat{y_t}-y_t}{y_t} \right|
\label{eq:9}
\end{equation}
Where $N$ is the number of forecasting data and it takes the value of 6 in our simulation. $y_t$ is real value of the electricity consumption, and the $\hat{y_t}$ is forecasting result.

The experiment¡¯s environment includes MATLAB R2016b, self-written MATLAB program and the computer with the Intel(R) Core(TM) i5-6500 3.2 GHz CPU, 16 GB RAM and Windows 7 professional system.

\subsection{Data set}
\label{Dataset}
\begin{figure}[htb]
\centerline{\includegraphics[scale=0.55]{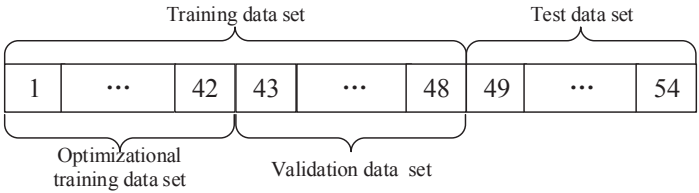}}
\caption{$~$The details of the sample data set.}
\label{dataset}
\end{figure}

The sample data set is provided by a Chinese energy supply company, which includes the monthly electricity consumption of a southern city in China from Jan. 2010 to Dec. 2018 and the monthly electricity consumption of the telecommunications and television industry in this city from Jan. 2010 to Dec. 2018. Since the meters of the users collect and report the electricity consumption every two months to the power management system, thus we only have 6 data points per year. Therefore, there are 54 data points in the sample data set. We select the data from Jan. 2010 to Dec. 2017 as training data set, i.e., the first 48 data points, and the data from Jan. 2018 to Dec. 2018 is used as the test data set, i.e., the last 6 points. Furthermore, we divide the training data set into optimizational training data set and validation data set, the optimizational training data set contains the first 42 data points of the training data set, and the validation data set contains the last 6 data points of the training data set. The optimizational training data set and the validation data set are used to conduct the FOA to search the optimal smoothing parameters for the Holt-Winters model, while the test data set is used to evaluate the performance of the finally forecasting results. The details of the sample data set is as Figure \ref{dataset}.

\subsection{The forecasting results of considered city}
\label{Province}
\begin{figure}[htb]
\centerline{\includegraphics[scale=0.4]{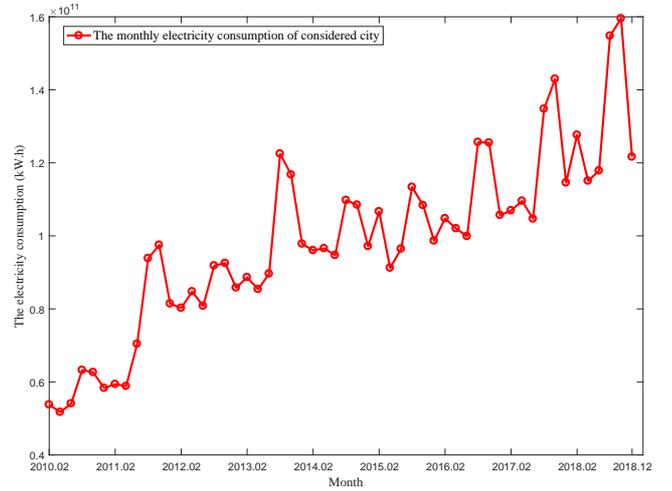}}
\caption{$~$The monthly electricity consumption of considered city.}
\label{ProvinceMES}
\end{figure}

The real monthly electricity consumption data of the considered city from Jan. 2010 to Dec. 2018 shows in Figure \ref{ProvinceMES}, and the forecasting process is as follows.

\begin{itemize}
  \item[(1)] Data set processing: According to the method presented in \ref{Dataset}, we get the training data set and test data.
  \item[(2)] Forecasting: Implement the process of FOA-MHW model which described in \ref{FOA-HW}. Specifically, initialize $maxgen$ and $sizepop$ by 20 and 50, respectively, $FX^\eta$ and $FY^\eta$ follow an independent uniform distribution of $[5, 10]$, $\eta\in{\alpha, \beta, \gamma}$, and the initial locations of these fruit fly swarms follow an independent uniform distribution over $[0, 1]$, i.e., $x_0^\eta, y_0^\eta \in[0,1],\eta\in{\alpha, \beta, \gamma}$.

  \item[(3)] Comparative analysis: We implement the seasonal index model, the HW model with default parameters and GASVR model in the same data set, where the smoothing parameters of Holt-winters exponential smoothing method is set to a default parameters, $\alpha = 0.2$, $\beta = 0.1$ and $\gamma = 0.6$ \cite{ForeastBook}.
\end{itemize}
\begin{figure}[htb]
\centerline{\includegraphics[scale=0.55]{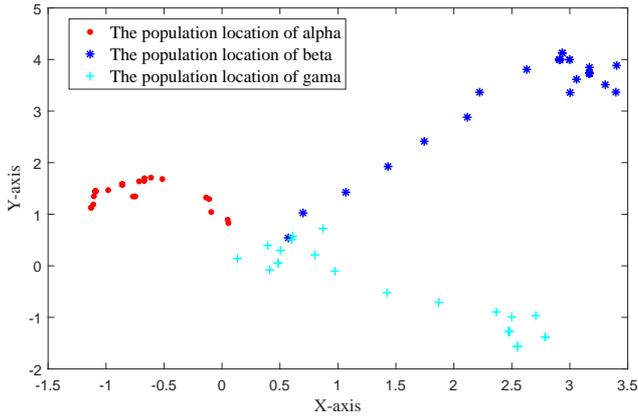}}
\caption{$~$The search route of fruit fly swarm for $\alpha$, $\beta$ and $\gamma$ for the considered city.}
\label{ProvinceRoute}
\end{figure}

\begin{figure}[htb]
\centerline{\includegraphics[scale=0.55]{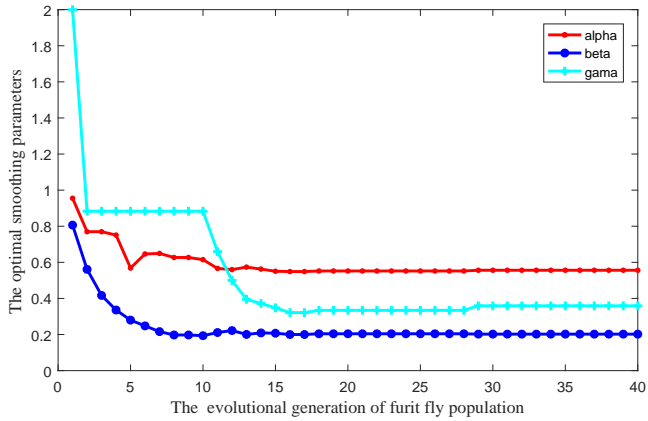}}
\caption{$~$The update process of optimal smoothing parameters in considered city.}
\label{ProvinceParameters}
\end{figure}
Therefore, by performing the forecasting process described above, we have the search route of the three fruit fly swarms as shown in Figure \ref{ProvinceRoute}, and the update process of the optimal smoothing parameters is shown in Figure \ref{ProvinceParameters}. One can note that the three fruit fly swarms can find the food quickly, i.e., the iteration of the smoothing parameters is finally converged. The optimal smoothing parameters selected by FOA are $\alpha = 0.5562$, $\beta = 0.2022$ and $\gamma = 0.3590$, however it's very difficult to obtain the optimal smoothing parameters by manually adjusting.

\begin{figure}[htbp]
\centerline{\includegraphics[scale=0.4]{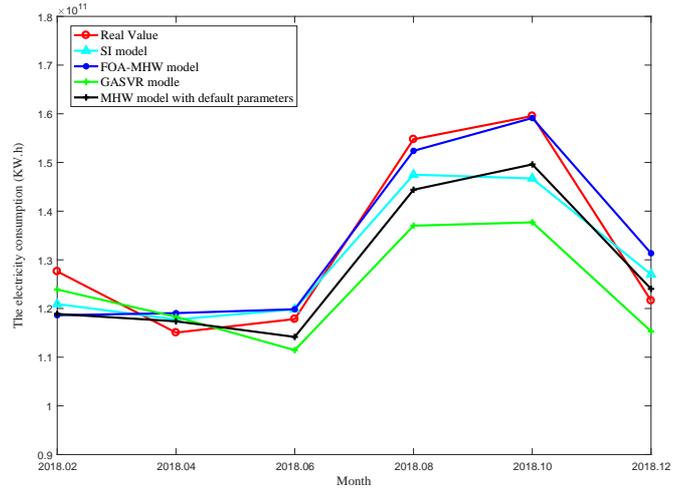}}
\caption{$~$The forecasting results of different models in considered city.}
\label{ProvinceResult}
\end{figure}

The forecasting results of different models are shown in Figure \ref{ProvinceResult}, meanwhile, we have compared the values of relative errors every month and the total MAPE for different models and they are listed in Table \ref{tab:MAPEProvince}. Firstly, we can note that, the proposed FOA-MHW model has the smallest MAPE, i.e., 3.65\%, and both the SI model and the MHW model with default parameters can achieve a better performance than the GASVR model. That is, though the time series of the monthly electricity consumption is highly randomness, leverage the optimal parameter selection of the fruit fly optimization algorithm (FOA), the FOA-MHW model can overcome this randomness at an acceptable level. However, since the order of the SI model and the HW model with default parameters are both lower than the FOA-MHW model, which then make them obtain worse performance than the FOA-MHW.

\begin{table}[htbp]
 \caption{\label{tab:MAPEProvince}The relative errors and MAPE of different models in considered city (\%)}
 \begin{tabular}{lllll}
  \toprule
  Month&SI model&MHW-default&GASVR&FOA-MHW\\
  \midrule
 Feb. 2018& 5.40& 7.00& 3.07& 7.20\\
 Apr. 2018& 2.16& 1.84& 2.63& 3.29\\
 July. 2018& 1.52& 3.28& 5.59& 1.54\\
 Aug. 2018& 4.82& 6.82& 11.58& 1.66\\
 Oct. 2018& 8.15& 6.35& 13.80& 0.39\\
 Dec. 2018& 4.26& 1.81& 5.35& 7.84\\
  MAPE&4.38& 4.52& 7.00& 3.65\\
  \bottomrule
 \end{tabular}
\end{table}

\subsection{The forecasting results of the telecommunications and television industry for the considered city}
\begin{figure}[htbp]
\centerline{\includegraphics[scale=0.55]{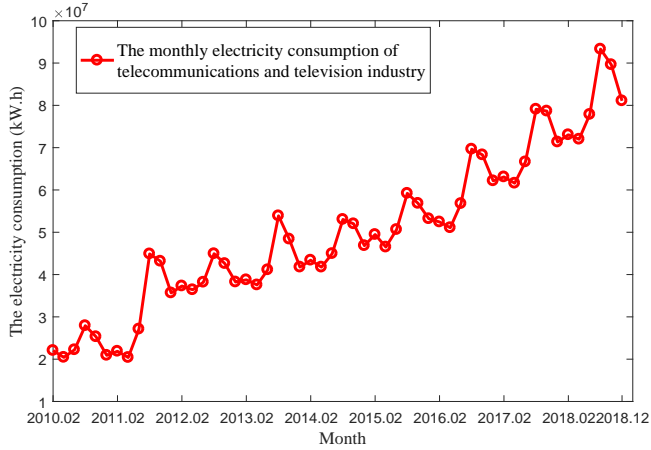}}
\caption{$~$The monthly electricity consumption of the telecommunications and television industry for the considered city.}
\label{IndustryMES}
\end{figure}

In order to further evaluate the performance of our model and understand its potential insight, we conduct another simulation but use a particular industry electricity consumption data, i.e., the telecommunications and television industry and which is shown in Figure \ref{IndustryMES}. One can note that, now, the time series of this industry have a better periodicity than the monthly electricity consumption of the considered city, i.e., the Figuie \ref{ProvinceMES}. The forecasting process and the parameter settings are the same as that described in section \ref{Province}.

\begin{figure}[htbp]
\centerline{\includegraphics[scale=0.55]{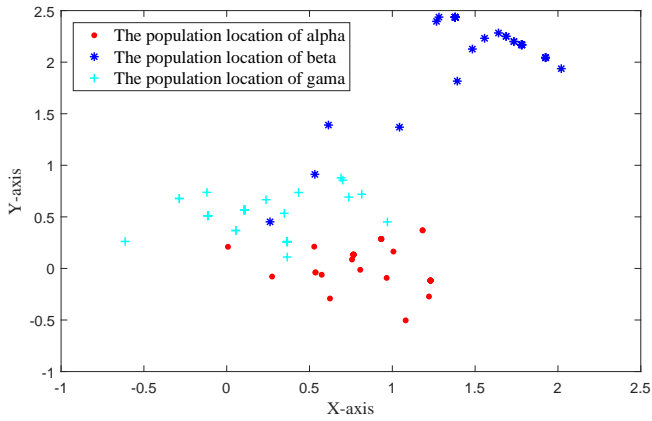}}
\caption{$~$The search route of fruit fly swarm for $\alpha$, $\beta$ and $\gamma$ of telecommunications and television industry.}
\label{IndustryRoute}
\end{figure}

\begin{figure}[htbp]
\centerline{\includegraphics[scale=0.55]{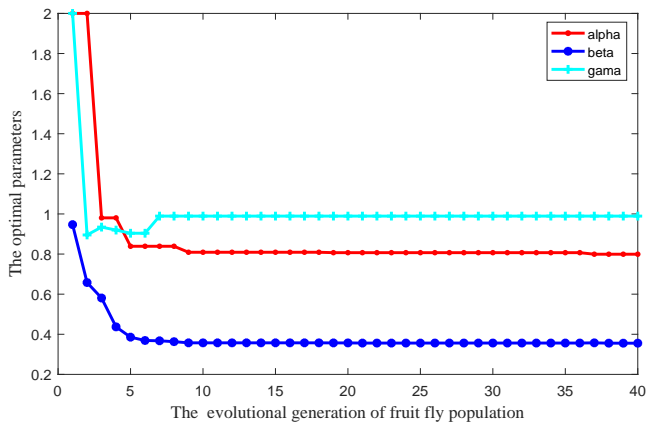}}
\caption{$~$The update process of optimal smoothing parameters of telecommunications and television industry.}
\label{IndustryParameters}
\end{figure}

The search route of the three fruit fly swarms are shown in Figure \ref{IndustryRoute}, and the update process of the optimal smoothing parameters is shown in Figure \ref{IndustryParameters}. Similarly, the three fruit fly swarms can find the optimal smoothing parameters very quickly. The optimal smoothing parameters selected by FOA is $\alpha = 0.7992$, $\beta = 0.3556$ and $\gamma = 0.9893$.

\begin{figure}[htbp]
\centerline{\includegraphics[scale=0.55]{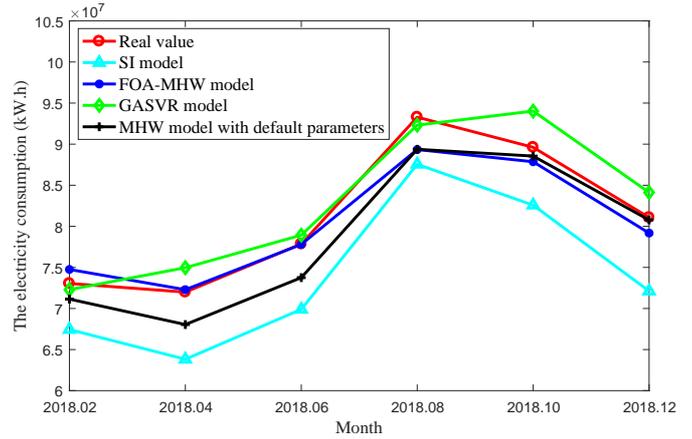}}
\caption{$~$The forecasting results of telecommunications and television industry.}
\label{IndustryResult}
\end{figure}

From the Figure \ref{IndustryResult} and the Table \ref{MAPEIndustry}, we can observe that, the FOA-MHW model still obtains the best performance over the other benchmark models in this particular industry. In specifically, the MAPE of our FOA-MHW model is 1.89\%, and the MAPE of the GASVR model and the SI model are 2.71\% and 9.05\%, respectively. And this result confirms that the FOA-MHW model can always get better performance than these three models for different scenarios.
\begin{table}[htbp]
 \caption{\label{MAPEIndustry}The relative errors and MAPE of different models in telecommunications and television industry(\%)}
 \begin{tabular}{lllll}
  \toprule
  Month&SI model&MHW-default&GASVR&FOA-MHW\\
  \midrule
 Feb. 2018& 7.68& 2.60& 1.04& 2.33\\
 Apr. 2018& 11.35& 5.48& 4.13& 0.41\\
 July. 2018& 10.22& 5.26& 1.38& 0.06\\
 Aug. 2018& 6.15& 4.19& 1.04& 4.23\\
 Oct. 2018& 7.86& 1.19& 4.92& 1.96\\
 Dec. 2018& 11.04& 0.37& 3.76& 2.34\\
  MAPE&9.05& 3.18& 2.71& 1.89\\
  \bottomrule
 \end{tabular}
\end{table}

\begin{figure}[htbp]
\centerline{\includegraphics[scale=0.55]{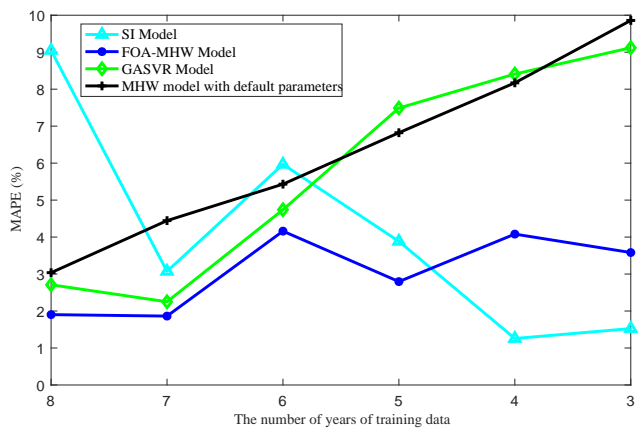}}
\caption{$~$The MAPE of different models in different lengthes of training data in telecommunications and television industry.}
\label{IndustryYear}
\end{figure}

At last, in order to evaluate the performance of our model in the case of small sample data set, the monthly electricity consumption with different lengthes of the training data set are forecasted. We use at least 3 years training data, i.e. 18 data points, and no more than 8 years training data, i.e. 48 data points. The MAPE of different models are shown in Figure \ref{IndustryYear}, we can find that, with the decreasing of the length of training data sets, the MAPE of the GA-SVR model and MHW model with default parameters increasing. In more detail, the MAPE of HW model with default parameters and the GASVR model increase linearly as the training data decreases, but the MAPE of our FOA-MHW model increase little as the training data decreases, and the MAPE of FOA-MHW model is 3.58\% even though there are only 3 years training data. One can note that, the MAPE of SI model is smallest when the training data less than 5 years, the reason is that the SI model can achieve a perfect performance when the data set has a regular periodicity, and the data from 2014 to 2018 is quite periodic in Figure \ref{IndustryMES}, so the SI model can get a best result, but the data from 2010-2013 isn't periodic enough in Figure \ref{IndustryMES}, which make the result of SI model is worse than other model. Besides, the MAPE of FOA-MHW model is smaller than the HW model with default parameters and the GASVR model all the time. The reason is that the machine learning method GASVR model need a lot of training data to get a good result, but the FOA-MHW model has a smaller data requirement. Thus, the FOA-MHW model applies to the situations of little training data available, where the machine learning method can't deal with.

\section{Conclusion}
\label{Conclusion}
In this paper, we deal with the problem of accurate monthly electricity consumption forecasting. In more detail, a new hybrid forecasting model named FOA-MHW is proposed herein. And the original idea is that, the fruit fly optimization algorithm is adopted for the parameters selection of the Holt-Winters exponential smoothing method. Based on the collected real electricity consumption data, the performance of the proposed model are testified over different scenarios. From which, we can conclude that, our proposed FOA-MHW model always achieve the best performance over other benchmark models, i.e., SI model, GASVR model and the MHW model with default parameters. In particular, our further experiments indicate that the proposed FOA-MHW model can obtain an excellent forecasting performance even with only a small number of sample data, e.g., the MAPE of FOA-MHW model is 3.58\% even though there are only 3 years training data. And now, our proposed FOA-MHW model has been deployed in a Chinese energy supply company to help them to forecast the monthly electricity consumption and its performance is well proved.




\setcounter{secnumdepth}{0} %
\section{Reference}
\bibliography{bibfile}
\end{document}